\documentclass[11pt,twoside]{article}
\usepackage{./asp2014}

\aspSuppressVolSlug
\resetcounters

\def\ab{Astrophys. Bull.}
\def\alet{Astron. Lett.}
\def\azh{Astron. Rep.}

\begin{document}

\title{Supergiants transiting towards white dwarfs}
\author{Valentina G. Klochkova,  
\affil{Special Astrophysical Observatory RAS, Nizhnij Arkhyz, Karachai-Cherkessia, Russia; email:{valenta@sao.ru}}}

\paperauthor{Valentina~Klochkova}{valenta@sao.ru}{ORCID_Or_Blank}{Special Astrophysical Observatory RAS}{Astrospectrosopy laboratory}{Nizhnij Arkhyz}{Karachai-Cherkessia}{369167}{Russia}

\begin{abstract}
Observational manifestations of far evolved stars at the asymptotic giants branch and their nearest descendants 
are briefly considered. Main results of their chemical composition determinations  based on long term high 
resolution spectroscopy at the 6-m telescope are also briefly summed up. A new kind of peculiarity of 
optical spectra is found and discussed: splitting or asymmetry of strongest absorptions in the optical 
spectra of selected post-AGB stars with envelopes and atmospheres enriched in carbon and s-process heavy metals. 
\end{abstract}

\section{Introduction}\label{intro}

Asymptotic giants branch (AGB)  and post-AGB stars are the principal suppliers of  heavy metals and important
suppliers of carbon and nitrogen to the interstellar medium. 
It is currently  believed that about  half of metals heavier than iron are synthesized in the s-process in deep 
layers of AGB--stars with initial masses  $ \le 3\div 4 \mathcal{M}_{\odot}$. 
Newly synthesized nuclei are moved to surface levels of the star by mixing  (the socalled third mixing, TDU) 
and, under the influence of the stellar wind, into the circumstellar medium. A detailed description 
of these processes and additional references can be found in reviews by \citet{Herwig} and \citet{Kappeler}. 
By supplying heavy metals 
and carbon to the interstellar  medium, AGB stars participate in the chemical evolution of their galaxies as a whole. 
However, although observations have enabled the identification of numerous evolved stars with carbon-enriched 
envelopes, the ejection of heavy metals into the circumstellar medium has not been directly detected.

On the Hertzsprung--Russell diagram stars with initial mass approximately $2\div 8 \mathcal{M}_{\odot}$  undergoing 
the short-lived  protoplanetary nebulae (PPN) stage evolve from the AGB toward the planetary nebula (PN)  
stage at almost constant luminosity,  getting increasingly hotter in the process. These descendants of AGB stars  are 
low-mass cores with typical masses of $0.6\,\mathcal{M}_{\odot}$  surrounded by an extended and often structured 
gaseous  envelope, which formed as a result of substantial mass loss by the star during the preceding evolutionary stages. 
The internal structure  of an AGB--star is very specific. A degenerate C/O core is surrounded by thin helium and hydrogen 
shells nuclear processing in which  become  in turn  active. Thus  final phase of AGB evolution is characterized  
by recurrent thermal pulses in helium shell.  Each thermal pulse  causes an increase in convection and dredge-up of 
fresh synthesized carbon and heavy metals to surface stellar layers. In the case of more massive stars with initial masses  
$3 \div 8 \mathcal{M}_{\odot}$  hot-bottom burning  (HBB) provides a synthesis of lower-mass chemical elements from 
Ne to Si (\citet{Herwig}).

\section{Observational data and main results}\label{data}

We have carried out numerous spectroscopic observations of supergiants with IR excesses using the 6-m telescope 
of the Special Astrophysical Observatory  over the past two decades. The initial aim of the program was to 
determine the fundamental  parameters of the program stars and to study chemical abundance anomalies in their 
atmospheres.  The combined information we have obtained can be used to reliably identify a star evolutionary stage. 
In the course of this program, we recognized the need for additional studies aimed at identifying spectral 
peculiarities and time variations of the spectral features. Studies of the velocity fields in the atmospheres 
and envelopes of the program stars were also required.

Spectral data were mainly  obtained at the Nasmyth focus  of the 6-meter telescope with the NES spectrograph 
developed by \citet{NES} using a 2048$\times$2048-pixel or 2048$\times$4096-pixel CCD chip and an image slicer, 
providing a spectroscopic resolving power of R\,=\,60\,000 within a  wide wavelength range to be observed simultaneously. 
The spectra of faint stars (the optical counterparts of the IR sources IRAS\,04296+3429 and 20000+3239) were
obtained with the PFES echelle spectrograph  designed by \citet{PFES} at the primary focus of the 6-m telescope. With
a 1024$\times$1024-pixel CCD chip, this spectrograph provides a resolving power of R\,=\,15\,000. Details of
our spectrophotometric and position measurements of the spectra are described by \citet{K2014}.

\begin{table}[!ht]
\smallskip
\begin{center}
{\small
\caption{Characteristics of circumstellar envelopes of the IR sources identified with the program post-AGB stars.
   The morphological type of the circumstellar envelope is given in accordance with \citet{Ueta,Sahai,Siodmiak,Lagadec}. 
   The velocity  $V_{\rm exp}$  was determined from the positions of CO and C$_2$ bands and the circumstellar components 
   of the Ba\,{\sc ii} lines. The appearance of the C$_2$ Swan bands (emission or absorption) is indicated in parantheses in 
   the fourth column.} 
\medskip
\begin{tabular}{l l  l  l  l }
\hline
IR source  &   Envelope     & \multicolumn{3}{c}{$V_{\rm exp}$, km/s} \\  
\cline{3-5}
 star      &    morphology          & \hspace{0.5cm} CO   & \hspace{0.5cm}  C$_2$ &  Ba\,{\sc ii} \\
\noalign{\smallskip}
 \hline
IRAS\,04296+3429& bipolar\,+    &10.8$^1$  & 7.7\,(abs) $^6$ & \\ 
CGCS\,6080 & halo\,+\,bar         &                           & 12\,(emis)$^7$  &   \\ 
\noalign{\smallskip}
\hline
IRAS\,07134+1005& elongated halo& 10.2$^1$      & 8.3\,(abs)$^6$ & \\  
CY\,CMi    &                    &                           & 11\,(emis)$^8$    & \\ 
\noalign{\smallskip}
\hline
IRAS\,08005$-$2356& bipolar     & 100:$^2$          &43.7\,(abs)$^6$ & \\ 
V510\,Pup  &                    &                           & 42\,(abs)$^9$   & \\ 
\noalign{\smallskip}
\hline
IRAS\,19500$-$1709& bipolar     &17.2, 29.5$^1$ &    no                &20 \& 30$^{10}$ \\ 
V5112\,Sgr &                    &10, 30--40$^3$   &                           & \\     
\hline        
IRAS\,20000+3239& elongated halo& 12.0$^4$       & 12.8\,(abs)$^6$& \\ 
CGCS\,6857 &                    &                           &11.1\,(abs)$^{11}$ & \\ 
\noalign{\smallskip}
\hline        
RAFGL\,2688 &multipolar\,+  &17.9, 19.7$^5$         & 17.3\,(abs)$^6$&\\ 
V1610\,Cyg& +\,halo\,+\,arcs    &                           & 60\,(emis)$^{12}$   & \\ 
\noalign{\smallskip}
\hline          
IRAS\,22223+4327& halo\,+\,lobes&14--15$^5$        & 15.0\,(abs)$^6$& \\ 
V448\,Lac &                     &                           &15.2\,(emis)$^{13}$ &  \\ 
\noalign{\smallskip}
\hline          
IRAS\,22272+5435&elongated halo\,+ &9.1--9.2$^1$ &9.1\,(abs)$^6$  & \\ 
V354\,Lac & +\,arcs             &                           &10.8\,(abs) $^{14}$&10$^{15}$ \\ 
\noalign{\smallskip}
\hline          
IRAS\,23304+6147&quadrupole\,+ & 9.2--10.3$^1$  &13.9\,(abs)$^6$ & \\ 
CGCS\,6918 & +\,halo\,+\,arcs   &                           &15.5\,(emis)$^{16}$ &15.1$^{16}$ \\ 
\noalign{\smallskip}
\hline 
\multicolumn{5}{l}{\scriptsize References: 1--\citep{Hrivnak},  2-- \citep{Hu},  3 -- \citep{Bujar},} \\  
\multicolumn{5}{l}{\scriptsize 4 -- \citep{Omont}, 5 -- \citep{Loup}, 6 -- \citep{Bakk97}, } \\
\multicolumn{5}{l}{\scriptsize  7 -- \citep{04296}, 8 -- \citep{atlas},  9 -- \citep{08005},  }\\  
\multicolumn{5}{l}{\scriptsize 10 -- \citep{19500}, 11 -- \citep{20000},  12 --\citep{Egg1},  }\\   
\multicolumn{5}{l}{\scriptsize 13 -- \citep{22223},  14 -- \citep{V354Lac}, 15 -- \citep{V354Lacb}, }\\
\multicolumn{5}{l}{\scriptsize 16 -- \citep{23304b}.}\\
\end{tabular}
\label{PPN}
}
\end{center}
\end{table} 

The primary conclusion of the analysis of the properties of the supergiants with infrared excess studied so 
far is that the available sample of these objects is inhomogeneous \citep{K2014}. These objects are found  
to exhibit a great  variety of peculiarities in the optical spectra of their central stars, the chemical composition 
of their atmospheres  and envelopes, and the morphology and kinematic state of their circumstellar envelopes. 
We should point out, in particular, that supergiants with infrared excess include RV\,Tau type variables. 
These variable stars with near-infrared excesses undergo the  post--AGB evolutionary stage, and, as it was shown 
by \citet{Winck2007}, most of them  are binaries. 

Besides some stars of the program were identified as  belonging  to  massive  stars far evolved. Most known objects  
with unclear status are high luminous stars V1302\,Aql (=\,IRC+10420)  and HD\,179821 (=\,IRAS\,19114+0002). 
Their properties (high luminosity, spectral peculiarities, presence of circumstellar envelopes) are similar 
to that observing for PPN. But thanks to the reinforced research during last decades (see \citet{kcp97,hds02,irc2016,sah2016} 
and numerous references  therein)  both stars are now considered to be the most unambiguous  massive stars with 
a highest mass loss rate which undergoes a short-term evolutionary transition from a red supergiant to a Wolf-Rayet star. 
Predecessors of the yellow hypergiants are massive (initial mass $\ge 20$ M$_{\odot}$)  and the most luminous stars, which lose 
a significant part of their mass after leaving main-sequence, become red supergiants, and later proceed to yellow supergiants.   

Quite opposite conclusions were obtained for the star  BD$-6^{\rm o}$1178  which was considered 
to be a candidate to PPN according to the observed excess of radiation in the 12--60\,$\mu$m wavelength region 
and its position on the IR colour--colour diagram. But \citet{05238} based on the high-resolution spectra  are shown 
for the first time that  BD$-6^{\rm o}$1178  is a spectroscopic binary (SB\,2). Their components have close spectral 
types and luminosity classes: F5\,IV--III and F3\,V.  The classification of BD$-6^{\rm o}$1178 
as a supergiant in the transition stage of becoming a planetary nebula does not confirm.  BD$-6^o$1178 
probably is a young pre-MS stars. Moreover it is possibly a member of the 1\,c subgroup of the Ori\,OB1 association.

The next important results of our study is the formation of a small homogeneous subsample of PPNe  whose atmospheres 
are enriched in carbon and heavy metals synthesized during the AGB evolution. All these objects are listed in Table\,\ref{PPN}.  
Their IR spectra  contain an  unidentified emission feature at 21\,$\mu$ \citep{Hrivnak2009,Kwok21} 
which arises  in the circumstellar  medium. 
The large excesses of carbon  and heavy metals in the atmospheres of these post-AGB stars in Table\,\ref{PPN} 
were found for the central stars of  IRAS\,04296+3429 \citep{04296}, 07134+1005  \citep{K1995}, 
23304+6147  \citep{23304}; 20000+3239 \citep{20000};  RAFGL 2688   \citep{Egg1}; and  for IRAS 22272+5435  \citep{V354Lac,V354Lacb}. 
The chemical abundances for five of the stars from Table\,\ref{PPN} were also determined by \cite{Winckel}. 

In addition to objects with carbon-enriched atmospheres, Table\,\ref{PPN} also contains the poorly studied IR source 
IRAS 08005$-$2356. Data on the atmospheric elemental abundances of its central star or the presence of the 21\,$\mu$ band
in its IR spectrum are not available. However, \citet{08005} consider IRAS 08005$-$2356 to be related to the other
sources in Table\,\ref{PPN}, because its optical spectrum contains Swan bands of C$_2$ and emissions of neutral hydrogen, 
with some metal lines displaying emission--absorption profiles, and its circumstellar envelope
was detected by \citet{Hu} in CO emission. Thus, we may suspect that IRAS 08005$-$2356 a full member of this small sample.

\begin{figure}[!ht]
\includegraphics[width=9cm,height=12cm,bb=40 70 550 770,clip]{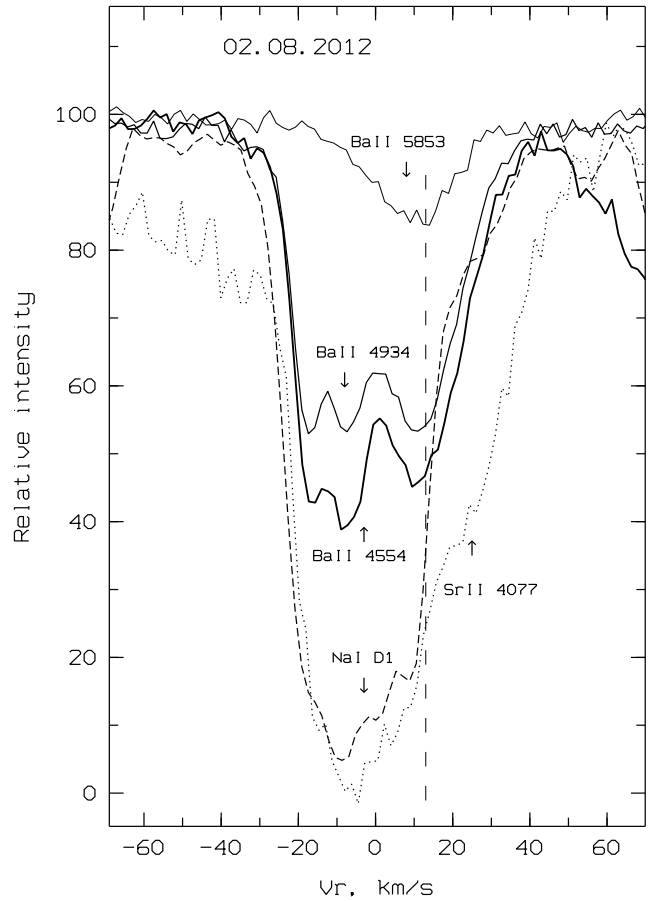}
\caption{Profiles of selected split lines in the spectrum of V5112\,Sgr taken in August, 02, 2012. The vertical dashed 
                     line indicates the systemic velocity Vsys\,=\,13\,km/s \citep{Bujar}.}
\label{V5112}
\end{figure}

We tried to consider  additional manifestations of circumstellar envelopes in the optical spectra
of post-AGB stars, concentrating on an analysis of this uniform sample of stars whose atmospheric 
chemical abundances have been shown in earlier former studies to have changed in the course of 
their evolution. Attempts to find an interconnection between the morphology of the envelope and the peculiarities of the 
chemical  composition of the central star revealed no strict correlation. It may nevertheless be concluded 
that objects  with enriched atmospheres (IRAS\,02229+6208, 04296+3429, 19500$-$1709, and 23304+6147) mostly 
have  structured  (bipolar, quadrupole, etc) envelopes. However, objects with enriched atmospheres also include IRAS\,05113+1347, 
whose envelope  even the HST failed to resolve, as well as  IRAS\,05341+0852 and 07134+1005 with envelopes 
in the form of  elongated haloes. 

One should also pay attention to IRAS\,19475+3119, which has a structured envelope,  whereas the atmosphere of 
its central star HD\,331319 is not enriched in either carbon or heavy metals. However, strong helium lines were 
found by  \citet{19475} in the spectrum of HD\,331319. The presence of these  lines in the spectrum  of a star 
with T$_{{\rm eff}}=7200$\,K results in a significant helium overabundance was detected  by \citet{19475} 
in its atmosphere. It should be mentioned here about detection of large lithium and sodium overabundance for 
the star  V2324\,Cyg (=\,IRAS\,20572+4919) (see more details \citet{20572}).  Results for these two stars permit us 
to propose that their initial mass was more than $3 \mathcal{M}_{\odot}$ and HBB and dredge-up  were effective 
at their AGB stage.

\section{Detection of a new type of peculiarities  in the optical spectra of post-AGB stars}\label{results}

Further, we concentrate on new properties of the optical spectra of a small subsample of post-AGB stars, 
from the Table\,\ref{PPN} whose circumstellar envelopes of these stars have complex morphologies, 
and are also enriched in carbon, manifest in the presence of C$_2$, C$_3$, CN, CO and other molecular bands. 
We analyze in detail main features of the optical spectra that form  in extended gas and dust envelopes and  
atmospheric radial velocity pattern measured on all  spectral features. The circumstellar envelopes of these 
stars contain extended halos, arcs, tori, etc. 
Some stellar envelopes display combinations of  these features, and they can possess bipolar or quadrupole 
circumstellar structures.  Optical spectra post-AGB supergiants differ from those of classical 
high-mass supergiants in the presence of molecular bands overlying the spectrum of the F--G supergiant, anomalies 
of the H\,{\sc i}, Na\,{\sc i}, and He\,{\sc i} absorption and emission profiles, and emission lines of some metals 
(for details see \citet{K2014}).  Moreover, all these spectral properties vary with time. 

In general, we see the following types of features in the optical spectra of post-AGB supergiants: 
1) symmetric absorption lines with low or moderate intensities, without obvious anomalies;
2) complex neutral-hydrogen line profiles containing time-variable absorption and emission components;
3) absorption or emission bands of molecules, often those containing carbon;  
4) shell components of the Na\,{\sc i} and K\,{\sc i} resonance lines; and 
5) narrow forbidden or permitted emission lines of metals formed in the envelope. 

The presence of features (2)--(5) is the main difference between the spectra of post-AGB stars and
high-mass supergiants.

Our many-year spectroscopic monitoring of post-AGB stars has enabled us to detect a previously unknown 
characteristic of the spectra of selected post-AGB stars: splitting (or asymmetry) of the strongest absorptions 
of heavy metals ions  (Sr\,{\sc ii}, Ba\,{\sc ii}, La\,{\sc ii}, Y\,{\sc ii}). 
This peculiarity has now been detected in the five stars presented in Table\,\ref{PPN}: CY\,CMi \citep{atlas}, 
V354\,Lac \citep{V354Lacb,V354Lac}, V448 Lac \citep{22223}, V5112\,Sgr \citep{19500}, and CGCS\,6918  
\citep{23304b}.
The strongest effect is for Ba\,{\sc ii} ions (see Fig.\,\ref{V5112}), whose lines with excitation
potentials of the lower levels $\chi_{low} \le 1$\,eV can be split  into two or three components. 
Infrared and  radio spectroscopy data are used to show that the stable individual  components of split 
absorptions are formed in structured circumstellar envelopes. Thus, this effect reveals efficient
enrichment of the envelope in heavy metals synthesized during the AGB evolution. We suggest that nature 
of the strong absorption profile (split or asymmetric, number of components) could be related to the 
morphology and kinematical and chemical properties of the envelope. 

The splitting of the profiles of the strongest absorptions of heavy metals detected in the spectra of the supergiants 
V5112\,Sgr, V354\,Lac, and CGCS\,6918, whose extended envelopes have complex structures, suggests that the formation of
a strong, structured envelope in the AGB stage is accompanied by the ejection of stellar nucleosynthesis
products into the circumstellar medium. Attempts to find a direct connection between the characteristics of 
the optical spectrum and the morphology of  the circumstellar medium are hindered by the fact that the observed 
structure of the envelope depends strongly on the inclination of the symmetry axis to the line of site, and 
also on the spectral and angular resolution of data and images used. 

Recall that the so called diffuse bands (DIBs) well known in the spectroscopy of the interstellar medium are 
present in the optical spectra of numerous post-AGB stars. But it should be here emphasized that in the spectrum 
of V5112\,Sgr we found several DIBs  whose accurate mean (from three spectra) radial velocity is in excellent agreement 
with the velocity derived from the circumstellar component of the Na\,I~D lines. This leads us to conclude that the
diffuse bands are  formed in the circumstellar envelope \citep{19500}.
  
We plan to continue in future our spectral research focusing on the fainter post-AGB stars with  strong  circumstellar 
envelopes. Complex morphology of envelopes suggests that these stars suffered in the past several  
events of mass loss  due to stellar winds of different rates. It means that we may expect manifestations of stellar 
nucleosynthesis in their atmospheres and circumstellar environment. For observations of such faint objects we need 
to improve the efficiency of our observational methods.

\acknowledgements 
I thank the organizers for the possibility to present this  review. I am grateful to all co-authors of my articles.
This study was accomplished with a financial support of the Russian Foundation for Basic Research  
in the framework of the projects No.\,11--02-00319\,a and 14--02--00291\,a.


\end{document}